\documentclass[12pt,onecolumn,aps,
               amsmath,amssymb,nofootinbib]{revtex4}      
\usepackage{graphicx}% Include figure files

\usepackage{graphicx,epsfig}

\usepackage{dcolumn}% Align table columns on decimal point
\usepackage{bm}% bold math

\newcommand{\Z}{{\mathbb Z}}

\begin{document}

\title{Universal Soft Terms in the MSSM on D-branes}

\author{Van E. Mayes}

\affiliation{Department of Chemistry, The University of Texas at Tyler,
Tyler, TX 75799}

\begin{abstract}
\begin{center}
{\bf ABSTRACT}
\end{center}
In Type II string vacua constructed from intersecting/magnetized D-branes, the supersymmetry-breaking
soft terms are genericaly non-universal. It is shown that universal supersymmetry-breaking soft terms 
may arise in a realistic MSSM constructed from intersecting/magnetized D-branes
in Type II string theory.  For the case of dilaton-dominated supersymmetry-breaking, it is shown
that the universal scalar mass and trilinear coupling are fixed such that 
$m_0=(1/2)m_{3/2}$ and $A_0 = - m_{1/2}$.  
In addition, soft terms where the universal scalar mass $m_0$ is much larger than the universal gaugino 
mass $m_{1/2}$ may be easily obtained within the model.
Finally, it is shown that the special dilaton and 
no-scale strict moduli boundary conditions, which are well-known in heterotic
string constructions, may also be obtained.    
\end{abstract}

\maketitle

\newpage

\section{Introduction}

Low-scale supersymmetry has been recognized for some time as the
most natural solution to the hierarchy problem.  In addition,
supersymmetry with R-parity imposed can provide a natural dark matter candidate
in the form of the Lightest Supersymmetric Partner (LSP), which is typically the lightest 
neutralino~\cite{Ellis:1982wr,Ellis:1983ew,Ellis:1983wd,Goldberg:1983nd}. 
Moreover, extending the
Standard Model (SM) to the Minimal Supersymmetric Standard Model (MSSM)
results in much-improved gauge coupling 
unification~\cite{Dimopoulos:1981yj, Ibanez:1981yh}.  Although the 
putative superpartners have yet to be observed, the recent discovery of a
Higgs boson with a mass near $125$~GeV~\cite{:2012gk,:2012gu} is consistent with the
upper bound on the Higgs mass in the MSSM, 
$m_H \lesssim 130$~GeV~\cite{Carena:2002es}.
 
Naively, one might have expected the superpartners to have already
been found based upon naturalness arguments which imply that the 
masses of the superpartners should be TeV-scale or lower.  
However direct searches from the ATLAS and CMS experiments at the 
Large Hadron Collider are pushing the mass limits on squarks and gluons above
the TeV-scale~\cite{:2012rz,Aad:2012hm,:2012mfa,Aad:2011ib,Chatrchyan:2011zy}.  
Furthermore, while the Higgs mass is below the MSSM 
upper bound, it is still somewhat larger than expected.  In order
to obtain such a large Higgs mass in the MSSM requires large 
radiative corrections from couplings to the top/stop quark
sector, implying muli-TeV
scale squark masses, and/or large values of tan$\beta$. 

Contrary to naive expectations, it is known that it is possible for
squarks and other scalars to have heavy multi-TeV masses while still solving
the hierarchy problem naturally, or at least by only introducing a small amount 
of fine-tuning.  
Perhaps the best known such scenario is that of the Hyperbolic Brand/Focus Point (HB/FP) 
supersymmetry~
\cite{Chan:1997bi,Feng:1999mn,Feng:1999zg,Baer:1995nq,Baer:1998sz,Chattopadhyay:2003xi,Akula:2011jx,Draper:2013cka}.  
FP superpartner spectra usually feature heavy multi-TeV scalars with lighter
gauginos.  The lightest neutralino in these spectra is typically of
mixed bino-higgsino composition, while the gluino is typically the heaviest of 
the gauginos and can have a mass up to a few TeV.  
In frameworks such as mSUGRA/CMSSM~\cite{Chamseddine:1982jx, Ohta:1982wn, Hall:1983iz},
focus point supersymmetry is realized in regions of the parameter
space occur where the universal scalar mass $m_0$ is much larger than
the universal gaugino mass, $m_{1/2}$.  It has been pointed out
that although spectra which lie on the focus point can solve the 
hierarchy problem with low fine-tuning of the electroweak scale,
this still requires a large amount of high-scale fine-tuning, at
least within the context of mSUGRA/CMSSM where $m_0 >> m_{1/2}$ is
unnatural as there is no {\it a priori} correlation of the high-scale 
parameters~\cite{Baer:2012mv}.

Although mSUGRA/CMSSM provides a simple and general framework
for studying the phenomenology of gravity-mediated supersymmetry
breaking, ultimately the supersymmetry breaking soft terms should
be determined within a specific model which provides a complete
description of physics at the Planck scale, such as string theory.
For example, in the context of Type II flux compactifications, soft terms 
of the form $m_0 >> m_{1/2}$ and $A_0 = -m_{1/2}$ may be induced by fluxes in Type IIB
string theory  with D$3$-branes~\cite{Camara:2003ku}.  
Soft terms of this form were studied
in ~\cite{Mayes:2013qmc} where it was shown to lead to focus-point regions of the
parameter space and where it is possible to obtain a $125$~GeV Higgs, satisfy
the WMAP9~\cite{Hinshaw:2012fq} and Planck~\cite{Ade:2013lta} 
results on the dark matter relic density as well
as all standard experimental constraints while 
maintaining low electroweak fine-tuning.

In the following, the possible sets of universal supersymmetry breaking
soft terms that may arise in an MSSM constructed from 
intersecting/magnetized D-branes in Type IIA/Type IIB string 
theory will be analyzed.  This model satisfies all global consistency conditions
and has many attractive features which
make it a suitable candidate for study.
These phenomenological features include three families of quarks and leptons,
a single pair of Higgs fields, automatic gauge coupling unification,
and exotics which are decoupled. From the low-energy effective
action of this model, it will be shown that the well-known special
dilaton solution may be obtained in the model
from the simplest set of possible F-terms, a result which should be generic to 
all models of this type.  It will then be shown
that there exist more general sets of universal soft terms in the model,
where supersymmetry-breaking is also dominated by the dilaton. For these
sets of soft terms, it will be shown that the universal scalar mass and
trilinear coupling are fixed so that $m_0=\frac{m_{3/2}}{2}$ and $A_0 = -m_{1/2}$,
where $m_{3/2}$ is the gravitino mass.
It will then be shown that universal 
soft terms where the universal scalar mass is much larger than the universal gaugino mass, 
$m_0 >> m_{1/2}$, may be obtained.
Finally, the no-scale strict moduli form of the soft
terms, $m_{1/2}=\frac{m_{3/2}}{2}$, $m_0 = A_0 = 0$ will be shown to be obtainable.  
The resulting phenomenology
will then be discussed.       
 
\section{The Model}

\begin{table}[t]
\caption{General spectrum for intersecting D6 branes at generic
angles, where $I_{aa'}=-2^{3-k}\prod_{i=1}^3(n_a^il_a^i)$ and
$I_{aO6}=2^{3-k}(-l_a^1l_a^2l_a^3
+l_a^1n_a^2n_a^3+n_a^1l_a^2n_a^3+n_a^1n_a^2l_a^3)$, where
$k = \beta_1+\beta_2+\beta_3$.  In addition,
${\cal M}$ is the multiplicity, and $a_S$ and $a_A$ denote
the symmetric and antisymmetric representations of
U($N_a/2$), respectively.}
\renewcommand{\arraystretch}{1.4}
\begin{center}
\begin{tabular}{|c|c|}
\hline {\bf Sector} & \phantom{more space inside this box}{\bf
Representation}
\phantom{more space inside this box} \\
\hline\hline
$aa$   & ${\rm U}(N_a/2)$ vector multiplet and 3 adjoint chiral
multiplets  \\
\hline $ab+ba$   & $ {\cal M}(\frac{N_a}{2},
\frac{\overline{N_b}}{2})=
I_{ab}=2^{-k}\prod_{i=1}^3(n_a^il_b^i-n_b^il_a^i)$ \\
\hline $ab'+b'a$ & $ {\cal M}(\frac{N_a}{2},
\frac{N_b}{2})=I_{ab'}=-2^{-k}\prod_{i=1}^3(n_{a}^il_b^i+n_b^il_a^i)$ \\
\hline $aa'+a'a$ &  ${\cal M} (a_S)= \frac 12 (I_{aa'} - \frac 12
I_{aO6})$~;~~ ${\cal M} (a_A)=
\frac 12 (I_{aa'} + \frac 12 I_{aO6}) $ \\
\hline
\end{tabular}
\end{center}
\label{spectrum}
\end{table}

Type II orientifold string compactifications with intersecting/magnetized 
D-branes  
have provided useful geometric tools with which the MSSM may
be engineered~\cite{Blumenhagen:2005mu,Blumenhagen:2006ci}.  
In the following, we shall work in the Type IIA 
picture with intersecting D6-branes, but it should be emphasized that
this model has a T-dual equivalent descriptioin in Type IIB with magnetized D-branes.
To briefly give an over view of the construction of such models, D6-branes in Type IIA fill
(3+1)-dimensional Minkowski spacetime and wrap 3-cycles in the
compactified manifold, such that a stack of $N$ branes generates a
gauge group U($N$) [or U($N/2$) in the case of $T^6/(\Z_2 \times
\Z_2)$] in its world volume.  On $T^6/(\Z_2 \times \Z_2)$, the
3-cycles are of the form~\cite{CSU}
\begin{equation}
\Pi_a = \prod_{i=1}^3(n_a^i[a_i] + 2^{-\beta_i} l_a^i[b_i]),
\end{equation}
where the integers $n_a^i$ and $l_a^i$ are the wrapping 
numbers around the basis cycles $[a_i]$ and $[b_i]$ of 
the $i$th two-torus, and $\beta_i=0$ for an untilted two-torus
while $\beta_i = 1$ for a tilted two-torus.    
In addition, we must introduce the orientifold images
of each D6-brane, which wraps a cycle given by
\begin{equation}
\Pi_a' = \prod_{i=1}^3(n_a^i[a_i] - 2^{-\beta_i} l_a^i[b_i]).
\end{equation}

In general, the 3-cycles wrapped by the stacks of D6-branes intersect
multiple times in the internal space, resulting
in a chiral fermion in the bifundamental representation localized at
the intersection between different stacks $a$ and $b$.  The multiplicity of such
fermions is then given by the number of times the 3-cycles intersect.
Each stack of D6-branes $a$ may 
intersect the orientifold images of other stacks $b'$, also resulting in fermions in
bifundamental representations.  Each stack may also intersect its own
image $a'$, resulting in chiral fermions in the symmetric and
antisymmetric representations.  The different types of representations
that may be obtained for each type of intersection and their
multiplicities are summarized in Table~\ref{spectrum}.  In addition, the
consistency of the model requires certain constraints to be satisfied,
namely, Ramond-Ramond (R-R) tadpole cancellation and the preservation 
of $\mathcal{N}=1$ supersymmetry.

\begin{table}[t]
\footnotesize
\renewcommand{\arraystretch}{1.0}
\caption{D6-brane configurations and intersection numbers for
the model on Type IIA $\mathbf{T}^6 / \Z_2 \times \Z_2$
orientifold. The complete gauge symmetry is $[U(4)_C \times U(2)_L
\times U(2)_R]_{\rm observable}\times [ USp(2)^4]_{\rm hidden}$, the SM
fermions and Higgs fields arise from the intersections on the
first two-torus, and the complex structure parameters are
$2\chi_1=6\chi_2=3\chi_3 =6$.}
\label{MI-Numbers}
\begin{center}
\begin{tabular}{|c||c|c||c|c|c|c|c|c|c|c|c|c|}
\hline
& \multicolumn{12}{c|}{$U(4)_C\times U(2)_L\times U(2)_R\times USp(2)^4$}\\
\hline \hline  & $N$ & $(n^1,l^1)\times (n^2,l^2)\times

(n^3,l^3)$ & $n_{S}$& $n_{A}$ & $b$ & $b'$ & $c$ & $c'$& 1 & 2 & 3 & 4 \\

\hline

    $a$&  8& $(0,-1)\times (1,1)\times (1,1)$ & 0 & 0  & 3 & 0 & -3 & 0 & 1 & -1 & 0 & 0\\

    $b$&  4& $(3,1)\times (1,0)\times (1,-1)$ & 2 & -2  & - & - & 0 & 0 & 0 & 1 & 0 & -3 \\

    $c$&  4& $(3,-1)\times (0,1)\times (1,-1)$ & -2 & 2  & - & - & - & - & -1 & 0 & 3 & 0\\

\hline

    1&   2& $(1,0)\times (1,0)\times (2,0)$ & \multicolumn{10}{c|}{$\chi_1=3,~
\chi_2=1,~\chi_3=2$}\\

    2&   2& $(1,0)\times (0,-1)\times (0,2)$ & \multicolumn{10}{c|}{$\beta^g_1=-3,~
\beta^g_2=-3$}\\

    3&   2& $(0,-1)\times (1,0)\times (0,2)$& \multicolumn{10}{c|}{$\beta^g_3=-3,~
\beta^g_4=-3$}\\

    4&   2& $(0,-1)\times (0,1)\times (2,0)$ & \multicolumn{10}{c|}{}\\

\hline

\end{tabular}

\end{center}

\end{table}

The set of D6 branes wrapping the cycles on a $T^6/(\Z_2 \times \Z_2)$
orientifold shown in Table~\ref{MI-Numbers} results in a
three-generation Pati-Salam model with additional hidden sectors.  The
full gauge symmetry of the model is given by $[{\rm U}(4)_C \times {\rm U}(2)_L \times {\rm
U}(2)_R]_{\rm observable} \times [ {\rm USp}(2)^4]_{\rm hidden}$,
with the matter
content shown in Table~\ref{Spectrum}.  As discussed in detail
in~\cite{Chen:2007px,Chen:2007zu}, with this configuration of D6 branes all R-R
tadpoles are canceled, K-theory constraints are satisfied, and
$\mathcal{N}=1$ supersymmetry is preserved.  

\begin{table}
[htb] \footnotesize
\renewcommand{\arraystretch}{1.0}
\caption{The chiral and vector-like superfields,
 and their quantum numbers
under the gauge symmetry $SU(4)_C\times SU(2)_L\times SU(2)_R
\times USp(2)_1 \times USp(2)_2 \times USp(2)_3 \times USp(2)_4$.}
\label{Spectrum}
\begin{center}
\begin{tabular}{|c||c||c|c|c||c|c|c|}\hline
 & Quantum Number
& $Q_4$ & $Q_{2L}$ & $Q_{2R}$  & Field \\
\hline\hline
$ab$ & $3 \times (4,\overline{2},1,1,1,1,1)$ & 1 & -1 & 0  & $F_L(Q_L, L_L)$\\
$ac$ & $3\times (\overline{4},1,2,1,1,1,1)$ & -1 & 0 & $1$   & $F_R(Q_R, L_R)$\\
$a1$ & $1\times (4,1,1,2,1,1,1)$ & $1$ & 0 & 0  & $X_{a1}$ \\
$a2$ & $1\times (\overline{4},1,1,1,2,1,1)$ & -1 & 0 & 0   & $X_{a2}$ \\
$b2$ & $1\times(1,2,1,1,2,1,1)$ & 0 & 1 & 0    & $X_{b2}$ \\
$b4$ & $3\times(1,\overline{2},1,1,1,1,2)$ & 0 & -1 & 0    & $X_{b4}^i$ \\
$c1$ & $1\times(1,1,\overline{2},2,1,1,1)$ & 0 & 0 & -1    & $X_{c1}$ \\
$c3$ & $3\times(1,1,2,1,1,2,1)$ & 0 & 0 & 1   &  $X_{c3}^i$ \\
$b_{S}$ & $2\times(1,3,1,1,1,1,1)$ & 0 & 2 & 0   &  $T_L^i$ \\
$b_{A}$ & $2\times(1,\overline{1},1,1,1,1,1)$ & 0 & -2 & 0   & $S_L^i$ \\
$c_{S}$ & $2\times(1,1,\overline{3},1,1,1,1)$ & 0 & 0 & -2   & $T_R^i$  \\
$c_{A}$ & $2\times(1,1,1,1,1,1,1)$ & 0 & 0 & 2   & $S_R^i$ \\
\hline\hline
$ab'$ & $3 \times (4,2,1,1,1,1,1)$ & 1 & 1 & 0  & \\
& $3 \times (\overline{4},\overline{2},1,1,1,1,1)$ & -1 & -1 & 0  & \\
\hline
$ac'$ & $3 \times (4,1,2,1,1,1,1)$ & 1 &  & 1  & $\Phi_i$ \\
& $3 \times (\overline{4}, 1, \overline{2},1,1,1,1)$ & -1 & 0 & -1  &
$\overline{\Phi}_i$\\
\hline
$bc$ & $6 \times (1,2,\overline{2},1,1,1,1)$ & 0 & 1 & -1   & $\eta_u^i$, $\eta_d^i$\\
& $6 \times (1,\overline{2},2,1,1,1,1)$ & 0 & -1 & 1   & \\
\hline
$bc'$ & $1 \times (1,2,2,1,1,1,1)$ & 0 & 1 & 1   & $H_u^i$, $H_d^i$\\
& $1 \times (1,\overline{2},\overline{2},1,1,1,1)$ & 0 & -1 & -1   & \\
\hline
\end{tabular}
\end{center}
\end{table}

The Pati-Salam gauge symmetry is broken to the SM in two 
steps~\cite{Cvetic:2004ui,Chen:2006gd}.  
First, the $a$ and $c$ stacks of D6-branes are split such that
$a \rightarrow a_1 + a_2$ and $c \rightarrow c_1 + c_2$, where
$N_{a_1}=6$, $N_{a_2}=2$, $N_{c_1}=2$, and $N_{c_2}=2$.  The process
of breaking the gauge symmetry via brane splitting corresponds
to assigning VEVs along flat directions to adjoint scalars associated
with each stack
that arise from the open-string moduli~\cite{Cvetic:2004ui}.   
After splitting the D6-branes,
the gauge symmetry of the observable sector is 
$SU(3)_C \times SU(2)_L \times U(1)_{I3R} \times U(1)_{B-L}$
where 
$U(1)_{I3R} = \frac{1}{2}(U(1)_{c1} - U(1)_{c2})$ 
and  
$U(1)_{B-L} = \frac{1}{3}(U(1)_{a1} - 3U(1)_{a2})$. 
The gauge symmetry may be further broken to that of the SM,
$SU(2)_C \times SU(2)_L  \times U(1)_Y$,
by assigning VEVs to vectorlike singlets in the $ac'$ sector,
where
$U(1)_Y = \frac{1}{6}\left[(U(1)_{a_1} - 3U(1)_{a_2} + 3(U(1)_{c_1} - U(1)_{c_2}\right]$.

\begin{table}
[htb] \footnotesize
\renewcommand{\arraystretch}{1.0}
\caption{The MSSM superfields,
and their quantum numbers
under the gauge symmetry $SU(3)_C\times SU(2)_L\times U(1)_Y$.}
\label{MSSMSpectrum}
\begin{center}
\begin{tabular}{|c||c||c|c|c||c|c|c|}\hline
 & Quantum Number
&  $Q_Y$  & Field \\
\hline\hline
$a_1b$ & $3 \times (3,\overline{2},1)$  & $\ 1/6$  & $Q_L$\\
$a_1b$ & $3 \times (1,\overline{2},1)$  & $-1/2$  & $L$\\
$a_1c_1$ & $3\times (\overline{3},1,1)$ & $\ 1/3$   & $U_R$\\
$a_1c_2$ & $3\times (\overline{3},1,1)$ & $-2/3$   & $D_R$\\
$a_2c_1$ & $3\times (1,1,1)$ & $\ 0$   & $N_R$\\
$a_2c_2$ & $3\times (1,1,1)$ & $-1$   & $E_R$\\
\hline
$bc_1'$ & $1 \times (1,2,1)$            &$\ 1/2$   & $H_u$\\
$bc_2'$ & $1 \times (1,2,1)$            & $-1/2$   & $H_d$ \\
\hline
\end{tabular}
\end{center}
\end{table}

The Higgs fields will be identified with the vectorlike fields
in the $bc'$ sector as opposed to the six vectorlie fields in 
the $bc$ sector as in previous studies of this model~\cite{Chen:2007zu}.
These multiplets are present since 
stacks $a$ and $c$ are parallel on the third two-torus, while 
stacks $a$ and $c'$ are parallel on the first torus.  It
will thus be assumed that stacks $a$ and $c$ are separated on the third
torus, while stacks $a$ and $c'$ are directly on top of one
another such that the Higgs fields in the $ac'$ sector 
remain massless while those from the $ac$ sector have
string-scale masses, and similarly for vectorlike matter in
the $ab'$ sector.  The additional fields in the model may become massive
as shown in~\cite{Chen:2007zu}.
Then, below the string scale the gauge symmetry and matter spectrum is that 
of the MSSM, as shown in Table~\ref{MSSMSpectrum}.

As shown in~\cite{Chen:2007zu}, this model has many desirable phenomenological
features.  In particular, the gauge couplings are automatically unified at
the string scale.  Furthermore, it was found that rank 3 Yukawa matrices for
quarks and charged leptons may be generated, and that it is possible to obtain their
observed masses and mixings.  In additional all chiral and vector-like exotics may
be decoupled from the low-energy spectrum.  It should be pointed out that the Higgs fields in
previous studies have been identified with the six vectorlike fields in the 
$bc$ sector of the model, rather than the vectorlike field in the $bc'$ as is
the case in the present study.  As a result, the trilinear Yukawa couplings
for quarks and leptons are forbidden by global symmetries.  However, the 
Yukawa couplings may in principle be generated by D-brane instantons or by
quartic couplings involving singlet fields.

\begin{table}
[t] \footnotesize
\renewcommand{\arraystretch}{1.0}
\caption{The angles (in multiples of $\pi$) with respect to the orientifold plane 
made by the cycle wrapped by each stack of D-branes 
on each of the three two-tori.}
\label{Angles}
\begin{center}
\begin{tabular}{|c||c||c|c|}\hline
 & $\theta_1$
&  $\theta_2$  & $\theta_3$ \\
\hline\hline
$a$ & $-1/2$  & $\ 1/4$  & $ 1/4$\\
$b$ & $\ 1/4$  & $\ 0 $  & $-1/4$\\
$c$ & $-1/4$ & $\ 1/2$   & $-1/4$\\
\hline
\end{tabular}
\end{center}
\end{table}

\section{The $\mathcal{N}=1$ Low-energy Effective Action}

From the effective scalar potential it is
possible to study the stability ~\cite{Blumenhagen:2001te}, the
tree-level gauge couplings \cite{CLS1, Shiu:1998pa,Cremades:2002te}, 
gauge threshold corrections \cite{Lust:2003ky},
and gauge coupling unification \cite{Antoniadis:Blumen}.  The
effective Yukawa couplings \cite{Cremades:2003qj, Cvetic:2003ch},
matter field K\"ahler metric and soft-SUSY breaking terms have
also been investigated \cite{Kors:2003wf}.  A more detailed
discussion of the K\"ahler metric and string scattering of gauge,
matter, and moduli fields has been performed in
\cite{Lust:2004cx}. Although turning on Type IIB 3-form fluxes can
break supersymmetry from the closed string sector
~\cite{Cascales:2003zp, MS, CL, Cvetic:2005bn, Kumar:2005hf,
Chen:2005cf}, there are additional terms in the superpotential
generated by the fluxes and there is currently no satisfactory
model which incorporates this. Thus, we do not consider this option
in the present work.  

The $\mathcal{N}=1$ supergravity action depends upon three
functions, the holomorphic gauge kinetic function, $f$, K\a"ahler
potential $K$, and the superpotential $W$.  Each of these will in
turn depend upon the moduli fields which describe the background
upon which the model is constructed. The holomorphic gauge kinetic
function for a D6-brane wrapping a calibrated three-cyce is given
by~\cite{Blumenhagen:2006ci}
\begin{equation}
f_P = \frac{1}{2\pi \ell_s^3}\left[e^{-\phi}\int_{\Pi_P} \mbox{Re}(e^{-i\theta_P}\Omega_3)-i\int_{\Pi_P}C_3\right].
\end{equation}
In terms of the three-cycle wrapped by the stack of branes, we have
\begin{equation}
\int_{\Pi_a}\Omega_3 = \frac{1}{4}\prod_{i=1}^3(n_a^iR_1^i + 2^{-\beta_i}il_a^iR_2^i).
\end{equation}
from which it follows that
\begin{eqnarray}
f_P &=&
\frac{1}{4\kappa_P}(n_P^1\,n_P^2\,n_P^3\,s-\frac{n_P^1\,l_P^2\,l_P^3\,u^1}{2^{(\beta_2+\beta_3)}}-\frac{n_P^2\,l_P^1\,l_P^3\,u^2}{2^{(\beta_1+\beta_3)}}-
\frac{n_P^3\,l_P^1\,l_P^2\,u^3}{2^{(\beta_1+\beta_2)}}),
\label{kingauagefun}
\end{eqnarray}
where $\kappa_P = 1$ for $SU(N_P)$ and $\kappa_P = 2$ for
$USp(2N_P)$ or $SO(2N_P)$ gauge groups and where we use the $s$ and
$u$ moduli in the supergravity basis.  In the string theory basis,
we have the dilaton $S$, three K\"ahler moduli $T^i$, and three
complex structure moduli $U^i$~\cite{Lust:2004cx}. These are related to the
corresponding moduli in the supergravity basis by
\begin{eqnarray}
\mathrm{Re}\,(s)& =&
\frac{e^{-{\phi}_4}}{2\pi}\,\left(\frac{\sqrt{\mathrm{Im}\,U^{1}\,
\mathrm{Im}\,U^{2}\,\mathrm{Im}\,U^3}}{|U^1U^2U^3|}\right)
\nonumber \\
\mathrm{Re}\,(u^j)& =&
\frac{e^{-{\phi}_4}}{2\pi}\left(\sqrt{\frac{\mathrm{Im}\,U^{j}}
{\mathrm{Im}\,U^{k}\,\mathrm{Im}\,U^l}}\right)\;
\left|\frac{U^k\,U^l}{U^j}\right| \qquad (j,k,l)=(\overline{1,2,3})
\nonumber \\
\mathrm{Re}(t^j)&=&\frac{i\alpha'}{T^j} \label{idb:eq:moduli}
\end{eqnarray}
and $\phi_4$ is the four-dimensional dilaton.
To second order in the string matter fields, the K\a"ahler potential is given by
\begin{eqnarray}
K(M,\bar{M},C,\bar{C}) = \hat{K}(M,\bar{M}) + \sum_{\mbox{untwisted}~i,j} \tilde{K}_{C_i \bar{C}_j}(M,\bar{M})C_i \bar{C}_j + \\ \nonumber \sum_{\mbox{twisted}~\theta} \tilde{K}_{C_{\theta} \bar{C}_{\theta}}(M,\bar{M})C_{\theta}\bar{C}_\theta.
\end{eqnarray}
The untwisted moduli $C_i$, $\bar{C}_j$ are light, non-chiral
scalars from the field theory point of view, associated with the
D-brane positions and Wilson lines.  In the following, it
will be assumed that these fields become massive via high-dimensional
operators.

For twisted moduli arising from strings stretching between stacks
$P$ and $Q$, we have $\sum_j\theta^j_{PQ}=0$, where $\theta^j_{PQ} =
\theta^j_Q - \theta^j_P$ is the angle between the cycles wrapped
by the stacks of branes $P$ and $Q$ on the $j^{th}$ torus
respectively. Then, for the K\a"ahler metric in Type IIA theory we find
the following two cases:

\begin{itemize}

\item $\theta^j_{PQ}<0$, $\theta^k_{PQ}>0$, $\theta^l_{PQ}>0$

\begin{eqnarray}
\tilde{K}_{PQ} &=& e^{\phi_4} e^{\gamma_E (2-\sum_{j = 1}^3
\theta^j_{PQ}) }
\sqrt{\frac{\Gamma(\theta^j_{PQ})}{\Gamma(1+\theta^j_{PQ})}}
\sqrt{\frac{\Gamma(1-\theta^k_{PQ})}{\Gamma(\theta^k_{PQ})}}
\sqrt{\frac{\Gamma(1-\theta^l_{PQ})}{\Gamma(\theta^l_{PQ})}}
\nonumber \\ && (t^j + \bar{t}^j)^{\theta^j_{PQ}} (t^k +
\bar{t}^k)^{-1+\theta^k_{PQ}} (t^l +
\bar{t}^l)^{-1+\theta^l_{PQ}}.
\end{eqnarray}

\item $\theta^j_{PQ}<0$, $\theta^k_{PQ}<0$, $\theta^l_{PQ}>0$

\begin{eqnarray}
\tilde{K}_{PQ} &=& e^{\phi_4} e^{\gamma_E (2+\sum_{j = 1}^3
\theta^j_{PQ}) }
\sqrt{\frac{\Gamma(1+\theta^j_{PQ})}{\Gamma(-\theta^j_{PQ})}}
\sqrt{\frac{\Gamma(1+\theta^k_{PQ})}{\Gamma(-\theta^k_{PQ})}}
\sqrt{\frac{\Gamma(\theta^l_{PQ})}{\Gamma(1-\theta^l_{PQ})}}
\nonumber \\ && (t^j + \bar{t}^j)^{-1-\theta^j_{PQ}} (t^k +
\bar{t}^k)^{-1-\theta^k_{PQ}} (t^l + \bar{t}^l)^{-\theta^l_{PQ}}.
\end{eqnarray}

\end{itemize}

For branes which are parallel on at least one torus, giving rise
to non-chiral matter in bifundamental representations (for example,
the Higgs doublets which arise from the bc' sector where stacks b and c' are parallel
on the first torus), the K\a"ahler metric is
\begin{equation}
K_{higgs}=((s+\bar{s})(t^2+\bar{t}^2)(t^3+\bar{t}^3)(u^1+\bar{u}^1))^{-1/2}.
\label{nonchiralK}
\end{equation}
The superpotential is given by
\begin{equation}
W = \hat{W}+ \frac{1}{2}\mu_{\alpha\beta}(M)C^{\alpha}C^{\beta} + \frac{1}{6}Y_(M){\alpha\beta\gamma}C^{\alpha\beta\gamma}+\cdots
\end{equation}
while the minimum of the F part of the tree-level supergravity
scalar potential $V$ is given by
\begin{equation}
V(M,\bar{M}) = e^G(G_M K^{MN} G_N -3) = (F^N K_{NM} F^M-3e^G),
\end{equation}
where
$G_M=\partial_M G$ and $K_{NM}=\partial_N \partial_M K$, $K^{MN}$
is inverse of $K_{NM}$, and the auxiliary fields $F^M$ are given
by
\begin{equation}
F^M=e^{G/2} K^{ML}G_L. \label{aux}
\end{equation}
Supersymmetry is broken when some of the F-terms of the hidden sector fields $M$
acquire VEVs. This then results in soft terms being generated in
the observable sector. For simplicity, it is assumed in this
analysis that the $D$-term does not contribute (see
\cite{Kawamura:1996ex}) to the SUSY breaking.  Then, the goldstino
is eaten by the gravitino via the superHiggs effect. The
gravitino then obtains a mass
\begin{equation}
m_{3/2}=e^{G/2}.
\end{equation}
The
normalized gaugino mass parameters, scalar mass-squared
parameters, and trilinear parameters respectively may be given in
terms of the K\a"ahler potential, the gauge kinetic function, and
the superpotential as
\begin{eqnarray}
M_P &=& \frac{1}{2\mbox{Re}f_P}(F^M\partial_M f_P), \\ \nonumber
m^2_{PQ} &=& (m^2_{3/2} + V_0) - \sum_{M,N}\bar{F}^{\bar{M}}F^N\partial_{\bar{M}}\partial_{N}log(\tilde{K}_{PQ}), \\ \nonumber
A_{PQR} &=& F^M\left[\hat{K}_M + \partial_M log(Y_{PQR}) - \partial_M log(\tilde{K}_{PQ}\tilde{K}_{QR}\tilde{K}_{RP})\right],
\label{softterms}
\end{eqnarray}
where $\tilde{K}_{QR}$ is the K\a"ahler metric appropriate for branes
which are parallel on at least one torus, i.e. involving
non-chiral matter.  In the present case, the Higgs
fields arise from vectorlike matter in the $bc'$ sector,
where the $b$ and $c'$ stacks are parallel on the first
two-torus.

\section{SUSY breaking via $u$-moduli and dilaton $s$}

We allow the dilaton $s$ to obtain a non-zero VEV as well as the $u$-moduli.
To do this, we parameterize the $F$-terms as
\begin{equation}
F^{u^i,s} = \sqrt{3}m_{3/2}[(s + \bar{s})\Theta_s e^{-i\gamma_s} + (u^i + \bar{u}^i)\Theta_i e^{-i\gamma_i}]
\end{equation}
The goldstino is included in the gravitino by $\Theta_S$ in $S$
field space, and $\Theta_i$ parameterize the goldstino direction
in $U^i$ space,  where $\sum (|\Theta_i^u|^2 + |\Theta_i^t|^2) + |\Theta_s|^2 =1$. The
goldstino angle $\Theta_s$ determines the degree to which SUSY
breaking is being dominated by the dilaton $s$ and/or complex
structure ($u^i$) and K\"ahler ($t^i$) moduli.

Then, the formula for
the gaugino mass associated with each stack can be expressed as
\begin{eqnarray}
M_P=\frac{-\sqrt{3}m_{3/2}}{4\mbox{Re} f_P}\left[\left(\sum_{j=1}^3
\mbox{Re} (u^j)\,\Theta_j\,
e^{-i\gamma_j}\,n^j_Pl^k_Pl^l_P2^{-(\beta_k +\beta_l)}\right) -
\Theta_s\mbox{Re}(s) e^{-i\gamma_0}n_P^1\,n_P^2\,n_P^3\, \right],
\\ \nonumber \qquad (j,k,l)=(\overline{1,2,3}).
\end{eqnarray}
The Bino mass parameter is a linear combination of the
gaugino mass for each stack, and the coefficients corresponding to
the linear combination of $U(1)$ factors define the hypercharge.

The trilinear parameters generalize as
\begin{eqnarray}
A_{PQR}&=&-\sqrt{3}m_{3/2}\sum_{j=0}^3 \left[
\Theta_je^{-i\gamma_j}\left(1+(\sum_{k=1}^3
 \xi_{PQ}^{k,j}\Psi(\theta^k_{PQ})-\frac{1}{4})+(\sum_{k=1}^3
 \xi_{RP}^{k,j}\Psi(\theta^k_{RP})-\frac{1}{4})
\right)\right] \nonumber \\
&&+\frac{\sqrt{3}}{2}m_{3/2}({\Theta}_{1}e^{-i{\gamma}_1} +
\Theta_s e^{-i{\gamma}_s}),
\end{eqnarray}
where $\Theta_0$ corresponds to $\Theta_s$  and there is a
contribution from the dilaton via the Higgs (1/2 BPS) 
K\a"ahler metric, which also gives an additional contribution to
the Higgs scalar mass-squared values:
\begin{equation}
m^2_H = m^2_{3/2}\left[1-\frac{3}{2}(\left|\Theta_1\right|^2+\left|\Theta_s\right|^2)\right].
\end{equation}
where $P$,$Q$, and $R$ label the stacks of branes whose mutual
intersections define the fields present in the corresponding
trilinear coupling and the angle differences are defined as
\begin{equation}
\theta_{PQ} = \theta_P - \theta_Q.
\end{equation}
We must be careful when dealing with cases where the angle difference is
negative.  Note for the present model, there is always either one
or two of the $\theta_{PQ}$ which are negative.  Let us define the
parameter
\begin{equation}
\eta_{PQ} = \mbox{sgn}(\prod_i \theta_{PQ}^i),
\end{equation}
such that $\eta_{PQ} = -1$ indicates that only one of the angle
differences are negative while $\eta_{PQ} = +1$ indicates that two
of the angle differences are negative.

Finally, the squark and slepton (1/4 BPS) scalar mass-squared
parameters are given as 
\begin{eqnarray}
m^2_{PQ}= m_{3/2}^2\left[1-3\sum_{m,n=0}^3
\Theta_m\Theta_ne^{-i(\gamma_m-\gamma_n)}\left(
\frac{{\delta}_{mn}}{4}+ \sum_{j=1}^3 (\xi^{j,m\bar
n}_{PQ}\Psi(\theta^j_{PQ})+
 \xi^{j,m}_{PQ}\xi^{j,\bar n}_{PQ}\Psi'(\theta^j_{PQ}))\right)
\right],
\end{eqnarray}
where we  include the $\Theta_s = \Theta_0$ in the sum. The
functions $\Psi(\theta_{PQ})$ and $\Psi'(\theta_{PQ})$ are given
by Eq.~(\ref{eqn:Psi1}) and Eq.~(\ref{eqn:Psi2}).  The
terms associated with the complex moduli in ${\xi}^{j,k}_{PQ}$ and
${\xi}^{j,k\bar{l}}_{PQ}$ are shown in 
Eq.~(\ref{idb:eq:dthdu}) and Eq.~(\ref{idb:eq:dth2du}),

The functions $\Psi(\theta_{PQ})=\frac{\partial \ln
(e^{-\phi_4}\tilde{K}_{PQ})}{\partial \theta_{PQ}}$ in the above
formulas defined for $\eta_{PQ}=-1$ are
\begin{eqnarray}
\mbox{if} \ \theta_{PQ} < 0&:& \\ \nonumber
\Psi(\theta^j_{PQ})&=&
-\gamma_E+\frac{1}{2}\frac{d}{d{\theta}^j_{PQ}}\,\ln{\Gamma(-\theta^j_{PQ})}-
\frac{1}{2}\frac{d}{d{\theta}^j_{PQ}}\,\ln{\Gamma(1+\theta^j_{PQ})}+\ln(t^j+\bar t^j)\\ \nonumber
\mbox{if} \ \theta_{PQ} > 0&:& \\ \nonumber
\Psi(\theta^j_{PQ})&=&
-\gamma_E+\frac{1}{2}\frac{d}{d{\theta}^j_{PQ}}\,\ln{\Gamma(1-\theta^j_{PQ})}-
\frac{1}{2}\frac{d}{d{\theta}^j_{PQ}}\,\ln{\Gamma(\theta^j_{PQ})}+\ln(t^j+\bar t^j),
\label{eqn:Psi1}
\end{eqnarray}
and for $\eta_{PQ}=+1$ are
\begin{eqnarray}
\mbox{if} \ \theta_{PQ} < 0&:& \\ \nonumber
\Psi(\theta^j_{PQ})&=&
\gamma_E+\frac{1}{2}\frac{d}{d{\theta}^j_{PQ}}\,\ln{\Gamma(1+\theta^j_{PQ})}-
\frac{1}{2}\frac{d}{d{\theta}^j_{PQ}}\,\ln{\Gamma(-\theta^j_{PQ})}-\ln(t^j+\bar t^j)\\ \nonumber
\mbox{if} \ \theta_{PQ} > 0&:& \\ \nonumber
\Psi(\theta^j_{PQ})&=&
\gamma_E+\frac{1}{2}\frac{d}{d{\theta}^j_{PQ}}\,\ln{\Gamma(\theta^j_{PQ})}-
\frac{1}{2}\frac{d}{d{\theta}^j_{PQ}}\,\ln{\Gamma(1-\theta^j_{PQ})}-\ln(t^j+\bar t^j).
\label{eqn:Psi2}
\end{eqnarray}
The function $\Psi'(\theta_{PQ})$ is just the derivative
\begin{eqnarray}
\Psi'(\theta^j_{PQ})&=&
\frac{d\Psi(\theta^j_{PQ})}{d \theta^j_{PQ}},
\label{eqn:Psip}
\end{eqnarray}
and ${\theta}^{j,k}_{PQ}$ and ${\theta}^{j,k\bar{l}}_{PQ}$ are
defined~\cite{Kane:2004hm} as

\begin{equation}
{\xi}^{j,k}_{PQ} \equiv (u^k+\bar u^k)\,\frac{\partial
\theta^j_{PQ}}{\partial u^k}= \left\{\begin{array}{cc}
 \left[-\frac{1}{4\pi}
 \sin(2\pi\theta^j)
 \right]^P_Q & \mbox{ when }j=k  \vspace*{0.6cm} \\
 \left[\frac{1}{4\pi}
\sin(2\pi\theta^j)
 \right]^P_Q & \mbox{ when }j\neq k,
\end{array}\right.\label{idb:eq:dthdu}
\end{equation}

\begin{equation}
{\xi}^{j,k\bar{l}}_{PQ} \equiv (u^k+\bar
u^k)(u^l+\bar u^l)\,\frac{\partial^2 \theta^j_{PQ}}{\partial
u^k\partial\bar u^l}= \left\{\begin{array}{cc} \frac{1}{16\pi}
  \left[ \sin(4\pi\theta^j)+4\sin(2\pi\theta^j)
 \right]^P_Q &
   \mbox{when }j=k=l  \vspace*{0.6cm} \\
 \frac{1}{16\pi}  \left[
 \sin(4\pi\theta^j)-4\sin(2\pi\theta^j)
 \right]^P_Q &
   \mbox{when }j\neq k=l  \vspace*{0.6cm} \\
 -\frac{1}{16\pi}\left[
 \sin(4\pi\theta^j)
 \right]^P_Q &
   \mbox{ when }j=k\neq l\mbox{ or } j=l\neq k \vspace*{0.4cm} \\
 \frac{1}{16\pi}\left[
\sin(4\pi\theta^j)
 \right]^P_Q &
   \mbox{when }j\neq k\neq l\neq j.
\end{array}\right.\label{idb:eq:dth2du}
\end{equation}

The terms associated with the dilaton are given by
\begin{equation}
{\xi}^{j,s}_{PQ} \equiv (s+\bar s)\,\frac{\partial
\theta^j_{PQ}}{\partial s}=
 \left[-\frac{1}{4\pi}
 \sin(2\pi\theta^j),
 \right]^P_Q \label{idb:eq:dthdus2}
\end{equation}
\begin{equation}
{\xi}^{j,k\bar s}_{PQ} \equiv (u^k+\bar
u^k)(s+\bar s)\,\frac{\partial^2 \theta^j_{PQ}}{\partial
u^k\partial\bar s}= \left\{\begin{array}{cc}
  \frac{1}{16\pi}\left[\sin{4\pi\theta^j}\right]^P_Q & \mbox{when} \ j=k \vspace*{0.6cm} \\
  -\frac{1}{16\pi}\left[\sin{4\pi\theta^j}\right]^P_Q & \mbox{when} \ j\neq k, \vspace*{0.2cm}
\end{array}\right.\label{idb:eq:dth2duds}
\end{equation}
and
\begin{equation}
{\xi}^{j,s\bar s}_{PQ} \equiv (s+\bar
s)(s+\bar s)\,\frac{\partial^2 \theta^j_{PQ}}{\partial
s\partial\bar s}=
  \frac{1}{16\pi}\left[\sin{4\pi\theta^j} + 4\sin(2\pi\theta^j)\right]^P_Q,\label{idb:eq:dth2dss}
\end{equation}
where $k,l\neq s$. The $\Theta_i$ parameters are
constrained as $\sum_{i=1}^3 \Theta_i^2 + \Theta_s^2 = 1$.

\section{The Special Dilaton}

First, we consider the case where the goldstino angles and dilaton are all equal, namely
$\Theta_1 = \Theta_2 = \Theta_3 = \Theta_s = 1/2$.  In addition,
we set $\gamma_1 = \gamma_2 = \gamma_3 = \gamma_s = 0$.

For the gaugino mass associated with each stack of D-branes we have

\begin{eqnarray}
M_P=\frac{-\sqrt{3}m_{3/2}}{4\mbox{Re} f_P}\left[\left(\sum_{j=1}^3
\mbox{Re} (u^j)\Theta_j\,
n^j_Pl^k_Pl^l_P2^{-(\beta_k +\beta_l)}\right) -
\mbox{Re}(s)\Theta_s n_P^1\,n_P^2\,n_P^3\, \right],
\\ \nonumber \qquad (j,k,l)=(\overline{1,2,3}),
\end{eqnarray}
where the holomorphic gauge kinetic function is given by 
\begin{eqnarray}
f_P &=&
\frac{1}{4}(n_P^1\,n_P^2\,n_P^3\,s-\frac{n_P^1\,l_P^2\,l_P^3\,u^1}{2^{(\beta_2+\beta_3)}}-\frac{n_P^2\,l_P^1\,l_P^3\,u^2}{2^{(\beta_1+\beta_3)}}-
\frac{n_P^3\,l_P^1\,l_P^2\,u^3}{2^{(\beta_1+\beta_2)}}),
\label{kingauagefun1}
\end{eqnarray}
from which it follows that there is a universal gaugino mass associated with each stack of Dbranes:
\begin{eqnarray}
M_P = \frac{-\sqrt{3}m_{3/2}}{8\mbox{Re} f_P}\left(-4f_P\right) = \frac{\sqrt{3}}{2}m_{3/2}.
\end{eqnarray}

The $Q_Y$ holomorphic
gauge function is given by taking a linear combination of the
holomorphic gauge functions from all the stacks. Note
that we have absorbed a factor of $1/2$ in the definition of $Q_Y$
so that the electric charge is given by $Q_{em} = T_3 + Q_Y$. In
this way, it is found~\cite{Blumenhagen:2003jy} that
\begin{equation}
f_Y = \sum_P \left|c_P\right| f_P = \frac{1}{6}f_{a1} + \frac{1}{2}f_{a2} + \frac{1}{2}f_{c1} + \frac{1}{2}f_{c2},
\end{equation}
where the the coefficients $c_P$ correspond to the linear
combination of $U(1)$ factors
which define the hypercharge, $U(1)_Y = \sum c_P U(1)_P$.
The Gaugino mass for $U(1)_Y$ is a linear combination of the
gaugino mass for each stack,
\begin{eqnarray}
M_{\tilde{B}} &=& \frac{1}{f_Y}\sum_P \left|c_P\right| f_P M_P 
     = \frac{\sqrt{3}}{2}m_{3/2} \frac{1}{f_Y}\sum_P \left|c_P\right| f_P \\ \nonumber
    &=& \frac{\sqrt{3}}{2}m_{3/2}.
\end{eqnarray}
Thus, the gaugino masses are universal:
\begin{equation}
m_{1/2} \equiv M_{\tilde{g}} = M_{\tilde{W}}= M_{\tilde{B}} = \frac{\sqrt{3}}{2}m_{3/2}.
\end{equation}

The Higgs scalar masses are given by
\begin{equation}
m^2_H = m^2_{3/2}\left[1-\frac{3}{2}(\left|\Theta_1\right|^2+\left|\Theta_s\right|^2)\right].
\end{equation}
With $\Theta_1 = \Theta_2 = \Theta_3 = \Theta_s = \frac{1}{2}$, we have
\begin{eqnarray}
m_H = \frac{m_{3/2}}{2} = \frac{m_{1/2}}{\sqrt{3}}.
\end{eqnarray}

For the trilinear couplings, we have
\begin{eqnarray}
A_0 \equiv A_{abc} = &-&\frac{\sqrt{3}}{2}m_{3/2}\sum_{j=0}^{3}\left(1+\left(\sum_{k=1}^3 \xi^{k,j}_{ab}\psi(\theta^k_{ab}) -\frac{1}{4}\right) +  \left(\sum_{k=1}^3 \xi^{k,j}_{ca}\psi(\theta^k_{ca}) -\frac{1}{4}\right)\right) \\ \nonumber
          &+&\frac{\sqrt{3}}{2}m_{3/2}. \\ \nonumber
        = &-& \frac{\sqrt{3}}{2}m_{3/2}(\frac{4}{2} + \sum_{j=0}^{3}\left(\xi^{1,j}_{ab}\psi(\theta^1_{ab}) + \xi^{2,j}_{ab}\psi(\theta^2_{ab}) + \xi^{3,j}_{ab}\psi(\theta^3_{ab})\right) \\ \nonumber
          &+& \sum_{j=0}^{3}\left(\xi^{1,j}_{ca}\psi(\theta^1_{ca}) + \xi^{2,j}_{ab}\psi(\theta^2_{ca}) + \xi^{3,j}_{ca}\psi(\theta^3_{ca})\right)) + \frac{\sqrt{3}}{2}m_{3/2}.
\end{eqnarray}
Now,
\begin{eqnarray}
\sum_{j=0}^{3}\xi^{i,j}_{PQ}\psi(\theta^i_{PQ}) = 0,
\end{eqnarray}
therefore we have
\begin{eqnarray}
A_0 &=& -\sqrt{3}m_{3/2} + \frac{\sqrt{3}}{2}m_{3/2} 
        = -\frac{\sqrt{3}}{2}m_{3/2} \\ \nonumber
        &=& -m_{1/2}.
\end{eqnarray}

The scalar masses for squarks and sleptons are given by
\begin{eqnarray}
m^2_{PQ} = m^2_{3/2}\left[1-\frac{3}{4}\sum_{m,n=0}^3\left(\frac{\delta_{mn}}{4}+\sum_{j=1}^3 \xi^{j,mn}_{PQ}\psi(\theta^j_{PQ}) + \xi^{j,m}_{PQ}\xi^{j,n}_{PQ}\psi'(\theta^j_{PQ})\right)\right].
\end{eqnarray}
Now,
\begin{eqnarray}
\sum_{m,n=0}^3 \xi^{i,mn}_{PQ}\Psi(\theta^i_{PQ}) = 0, \ \ \ \mbox{and} \ \ \  
\sum_{m,n=0}^3 \xi^{i,m}_{PQ}\xi^{i,n}_{PQ}\psi'(\theta^i_{PQ}) = 0.
\end{eqnarray}
Thus, we find that the scalar masses for squarks and sleptons are universal:
\begin{eqnarray}
m_{PQ} = \frac{m_{3/2}}{2}.
\end{eqnarray}

In summary, taking all goldstino angles to equal yields universal soft terms of the form
\begin{eqnarray}
m_{1/2} = \frac{\sqrt{3}}{2} m_{3/2}, \ \ \ \ \  m_0 = \frac{m_{3/2}}{2}, \ \ \ \ \  A_0 = -m_{1/2}.
\end{eqnarray} 
It should be noted that this solution for the soft terms is more-or-less 
model independent and should be present for any Pati-Salam model of this
type constructed from intersecting/magnetized D-branes.

\section{General Dilaton-dominated Soft Terms}

We have seen in the previous section that if all of the goldstino angles are equal,
then the soft terms are universal and we obtain the well-known special dilaton solution.  
In the following, let us consider more general possibilities where universal 
soft terms may be obtained. 

For the present model, the complex structure moduli and dilaton in the field theory
basis are given by 
\begin{eqnarray}
\mbox{Re}(u^1) = 2.61, \ \ \ \ \ \ \mbox{Re}(u^2) = 7.83, \ \ \ \ \ \ \mbox{Re}(u^3) = 3.915, \ \ \ \ \ \ \mbox{Re}(s) = 1.305,
\end{eqnarray}
while the real part of the holomorphic gauge kinetic functions for each stack are given
by 
$f_a = f_b = f_c = 1.9575$ and $f_Y = 3.2625$
from which it follows that the MSSM gauge couplings are unified at the string scale,
$g^2_3 = g^2_2 = \frac{5}{3}g^2_1$~\cite{Chen:2007zu}.  A scan of some of the soft
terms for non-universal soft terms was made in~\cite{Chen:2007zu}, and some of the 
phenomenological consequences have been studied in~\cite{Maxin:2009ez,Maxin:2009qq}. 

The gaugino masses may be written in terms of the goldstino angles as
\begin{eqnarray}
&M_3 = \frac{\sqrt{3}m_{3/2}}{2}(\Theta_2 + \Theta_3),  \\ \nonumber
&M_2 = \frac{\sqrt{3}m_{3/2}}{2}(\Theta_2 + \Theta_s),  \\ \nonumber
&M_{\tilde{B}} = \frac{\sqrt{3}m_{3/2}}{10}(3\Theta_1 + 2\Theta_2 + 5\Theta_3).
\end{eqnarray}
From these expressions, it can be seen that setting $\Theta_{12} \equiv \Theta_1 = \Theta_2$ 
and $\Theta_{3s} \equiv \Theta_3 = \Theta_s$ results in universal gaugino masses of the form
\begin{equation}
m_{1/2} \equiv M_3 = M_2 = M_{\tilde{B}} = \frac{\sqrt{3}m_{3/2}}{2}(\Theta_{12} + \Theta_{3s}).
\end{equation} 

The trilinear A-term may also be written in terms of the goldstino angles as
\begin{eqnarray}
A_0 = A_{abc} &=& \frac{\sqrt{3}}{2}m_{3/2}(\Theta_2 + \Theta_3)  \\ \nonumber
&-&\frac{\sqrt{3}}{4\pi}m_{3/2}[\left(-\Theta_s - \Theta_1 + \Theta_2 + \Theta_3\right)\left(\Psi\left(-3/4\right) + \Psi\left(1/4\right)\right)
\\ \nonumber &+& \left(-\Theta_s + \Theta_1 - \Theta_2 + \Theta_3\right)\left(\Psi\left(1/4\right) + \Psi\left(1/4\right)\right)]]. 
\end{eqnarray}
Letting $\Theta_1 = \Theta_2$ and $\Theta_3 = \Theta_s$ as in the gaugino masses,
the universal trilinear coupling takes the same simple form as for the special dilaton:
\begin{equation}
A_0 = -\frac{\sqrt{3}}{2}m_{3/2}(\Theta_{12} + \Theta_{3s}) = -m_{1/2}. 
\end{equation}
However, in the 
present case the relationship between the gaugino mass and the gravitino mass 
may be different than for the special dilaton, depending upon the values assigned
to $\Theta_{12}$ and $\Theta_{3s}$.   This will have important consequences for
phenomenology, as shall be discussed later. 

Next, let us turn to the expressions for the scalar masses. As before, the
Higgs scalar masses are given by 
\begin{eqnarray}
m^2_H &=& m^2_{3/2}\left[1-\frac{3}{2}\left(\Theta_{1}^2 + \Theta_{s}^2\right)\right] 
      = m^2_{3/2}\left[1-\frac{3}{2}\left(\Theta_{12}^2 + \Theta_{3s}^2\right)\right] \\ \nonumber
      &=& \frac{m^2_{3/2}}{4},       
\end{eqnarray}
where $\Theta_{12}^2 + \Theta_{3s}^2 = 1/2$.  It should be noted that this result
is the same as for the special dilaton solution of the previous section. 

For $\Theta_1=\Theta_2$ and $\Theta_s=\Theta_3$, the scalar masses for squarks and sleptons may be written as
\begin{eqnarray}
m^2_{PQ} = \frac{m^2_{3/2}}{4} &-& \frac{3m^2_{3/2}}{16\pi}\Theta^2_{3s}\cdot\left[\mbox{sin}4\pi\theta^3+4\mbox{sin}2\pi\theta^3\right]^P_Q + \Theta^2_{12}\cdot\left[\mbox{sin}4\pi\theta^3-4\mbox{sin}2\pi\theta^3\right]^P_Q \\ \nonumber
         &+& 2\left[\mbox{sin}4\pi\theta^3\right]^P_Q\cdot(\Theta_{3s}^2 + \Theta_{12}^2 -4\Theta_{3s}\cdot\Theta_{12})\cdot\Psi\left(\theta^3_{PQ}\right). 
\end{eqnarray}
Inserting the appropriate angles as shown in Table~\ref{Angles}, this expression then becomes
\begin{eqnarray}
m^2_{PQ} = m^2_{3/2}\left[\frac{1}{4} -\frac{3}{\pi}\left(\Theta_{3s}^2 - \Theta_{12}^2\right)\cdot\Psi_{PQ}\left(1/2\right)\right],
\end{eqnarray} 
where 
\begin{eqnarray}
\Psi_{ab}\left(1/2\right)&=& -\gamma_E+\frac{1}{2}\frac{d}{d{\theta}^3}\,\ln{\Gamma(1/2)}-
\frac{1}{2}\frac{d}{d{\theta}^3}\,\ln{\Gamma(1/2)}+\ln(t^3+\bar t^3) \\ \nonumber
                     &=& -\gamma_E + \ln(t^3+\bar t^3),
\end{eqnarray}
and
\begin{eqnarray}
\Psi_{ac}\left(1/2\right) &=& \gamma_E+\frac{1}{2}\frac{d}{d{\theta}^3}\,\ln{\Gamma(1/2)}-
\frac{1}{2}\frac{d}{d{\theta}^3}\,\ln{\Gamma(1/2)}-\ln(t^3+\bar t^3) \\ \nonumber
                     &=& \gamma_E - \ln(t^3+\bar t^3).
\end{eqnarray}
Then, if we set $\ln(t^3+\bar t^3) = \gamma_E$ we then obtain a universal scalar mass given
by
\begin{eqnarray}
m_0 \equiv m_H = m_{ab} = m_{ac} = \frac{m_{3/2}}{2}.
\end{eqnarray}

In summary, setting $\Theta_{12} \equiv \Theta_1 = \Theta_2$, $\Theta_{3s} \equiv \Theta_3 = \Theta_s$, and
$\mbox{ln}(t^3+\bar t^3) = \gamma_E$ results in universal soft terms of the form:
\begin{eqnarray}
m_{1/2} = \frac{\sqrt{3}}{2}(\Theta_{12} + \Theta_{3s}), \ \ \ \ \ \ 
m_0     = \frac{m_{3/2}}{2}, \ \ \ \ \ \ 
A_0     = -m_{1/2}.
\end{eqnarray}
These soft terms appear to be a generalized form of the special dilaton solution.  
In particular, setting all goldstino angles equal results in precisely the special
dilaton.  However, in the present case one may obtain a different result for 
the gaugino mass for more general goldstino angles.

\section{No-scale Moduli-dominated Soft Terms}

Let us include the K\a"ahler moduli in the supersymmetry breaking by parameterizing 
the F-terms as
\begin{eqnarray}
&& F^s=\sqrt{3} m_{3/2} {\rm Re}(s) \Theta_s e^{-i\gamma_s},
\nonumber \\
&&F^{\{u,t\}^i} = \sqrt{3} m_{3/2}( {\rm Re}  ({u}^i) \Theta_i^u
e^{-i\gamma^u_i}+  {\rm Re} ({t}^i) \Theta_i^t
e^{-i\gamma_i^t}).
\end{eqnarray}
In the following we shall take $\Theta_{12} \equiv \Theta_1^u = \Theta_2^u$ and
$\Theta_{3s} \equiv \Theta_{3}^u = \Theta_s$ and set the CP violating phases to
zero.  In addition, we shall set $\mbox{ln}(t_3+\overline{t}_3) = \gamma_E$ in order 
to have a universal scalar mass for squarks and sleptons.  

Then, the gaugino masses take the same universal form as before:
\begin{eqnarray}
m_{1/2}  = \frac{\sqrt{3}m_{3/2}}{2}(\Theta_{12} + \Theta_{3s}).
\end{eqnarray} 
The Higgs scalar massses then become
\begin{eqnarray}
m^2_H = m^2_{3/2}\left[1-\frac{3}{2}\left(\Theta_{12}^2 + \Theta_{3s}^2 + (\Theta_{2}^t)^2 +(\Theta_{3}^t\right))^2\right], 
\end{eqnarray}
while the scalar masses for squarks and sleptons becomes
\begin{eqnarray}
m^2_{PQ} = m^2_{3/2}\left[1-\frac{3}{2}\left(\Theta^2_{12}+\Theta^2_{3s}\right) - 3\left(\frac{3}{4}(\Theta_1^t)^2 + \frac{3}{4}(\Theta_2^t)^2 + \frac{1}{2}(\Theta_3^t)^2\right)\right].
\end{eqnarray}

Assuming that the dependence of the soft terms on the Yukawa couplings through the K\a"ahler moduli 
may be ignored, the trilinear coupling becomes
\begin{eqnarray}
A_0 = -\frac{\sqrt{3}}{2}m_{3/2}(\Theta_{12} + \Theta_{3s}+ 2\Theta_1^t+ \Theta_2^t+ \Theta_3^t).  
\end{eqnarray}

If we take $\Theta_{12} = -\Theta_3^t = \frac{1}{\sqrt{3}}$, and $\Theta_{3s} = \Theta_1^t = \Theta_2^t = 0$, we obtain the 
no-scale strict moduli scenario:
\begin{eqnarray}
m_{1/2} = \frac{m_{3/2}}{2}, \ \ \ \ \ \  m_0 = 0, \ \ \ \ \ \ A_0 = 0.
\end{eqnarray}

Note that in this case, the supersymmetry breaking is dominated by the K\a"ahler moduli,
while the dilaton does not participate.  This is just what is expected for the no-scale
form and is referred to as the strict moduli-dominated scenario. 

\begin{figure}
  \centering
	\includegraphics[width=1.0\textwidth]{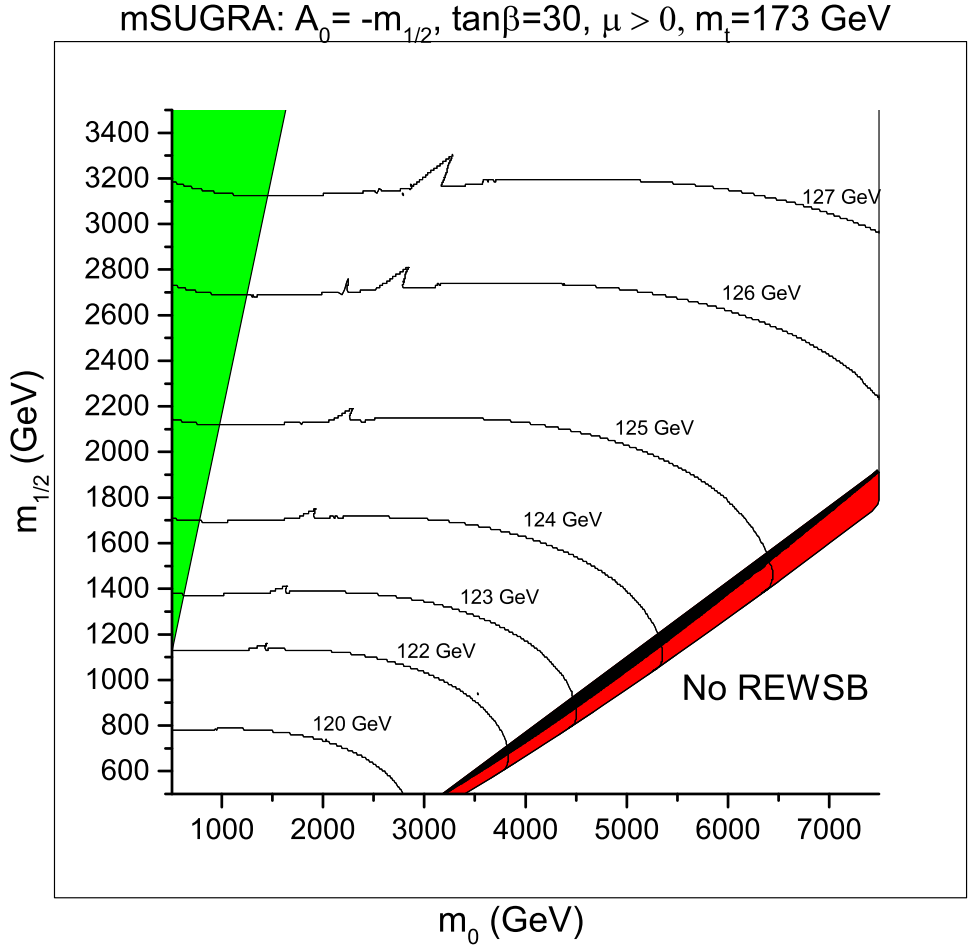}
		\caption{The mSUGRA $m_{1/2} \ vs.\ m_0$ plane with $A_0 = -m_{1/2}$, $\mu>0$, tan$\beta=30$, and $m_t=173$~GeV.  The region shaded in black indicates
		a relic density $0.105 \lesssim \Omega_{\chi^0} h^2 \lesssim 0.123$, the region shaded in red indicates $\Omega_{\chi^0} h^2 \lesssim 0.123$, while the region shaded in green has a charged LSP.  
		The black contour lines indicate the lightest CP-even Higgs mass.}
	\label{fig:mSUGRA_CountourPlanetb30}
\end{figure}

\begin{figure}
  \centering
	\includegraphics[width=1.0\textwidth]{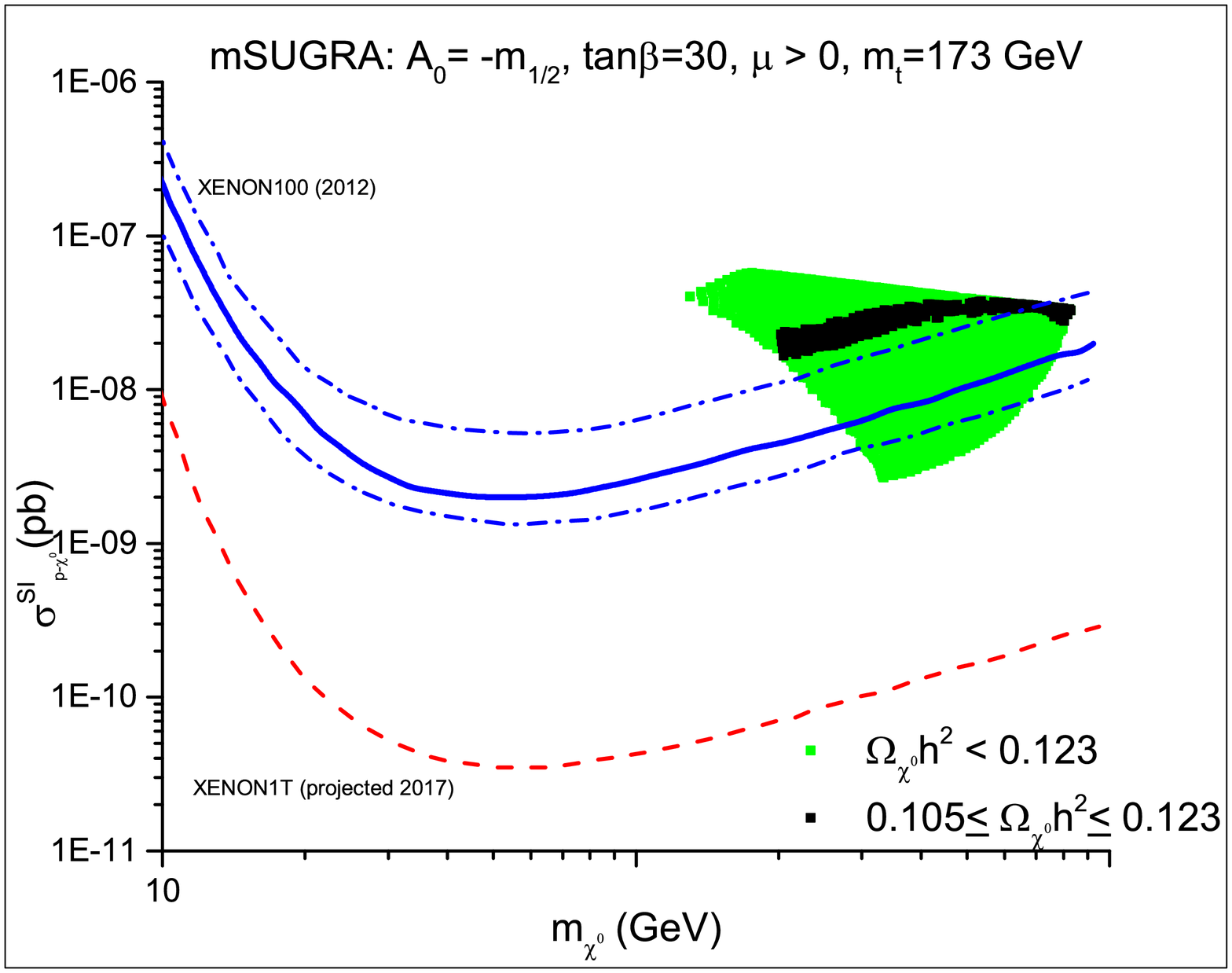}
		\caption{The spin-independent (SI) neutralino-proton direct detection cross-sections vs. neutralino mass for regions of the parameter space where $\Omega_{\chi^0}h^2\leq 0.123$. The region shaded
		in black indicates $0.105 \lesssim \Omega_{\chi^0}h^2 \lesssim 0.123$.  The upper limit on the cross-section obtained from the XENON100 experiment is shown in blue with the $\pm 2\sigma$ bounds shown as dashed curves, while the red dashed curved indicates the future reach of the XENON1T experiment.}
	\label{fig:DirectDetCrossSectionsvsNeutralinoMasstb30A}
\end{figure}

\section{Phenomenological Discussion}

In the previous sections, several different forms for the supersymmetry breaking
soft terms that may arise from a realistic intersecting/magnetized D-brane
model were discussed.  Two of these are well-known, namely the special dilaton form
and the no-scale strict moduli form which arise from dilaton-dominated and K\a"ahler
moduli dominated supersymmetry respectively.  The phenomenology of both of these scenarios
has been extensively explored, and neither of these two cases presently has a viable parameter 
space which can satisfy experimental constraints~\cite{Maxin:2009pr,Maxin:2008kp}. 

On the other hand, a different form for the soft terms was also explored,
which appears to be a generalized form of dilaton-dominated supersymmetry
breaking. 
In particular, it was found that if $\Theta_1 = \Theta_2 \equiv \Theta_{12}$ 
and $\Theta_3 = \Theta_s \equiv \Theta_{3s}$, the universal trilinear term
is always equal to the negative of the gaugino mass, $A_0 = -m_{1/2}$.
Furthermore, the universal scalar mass is given by 
\begin{eqnarray}
m_{1/2} = \frac{\sqrt{3}m_{3/2}}{2}(\Theta_{12} + \Theta_{3s}) 
        = \sqrt{3} m_0 (\Theta_{12} + \Theta_{3s}).
\end{eqnarray}
From this expression, it may been observed that $m_0$ is always larger
than the universal gaugino mass, $m_0 > m_{1/2}$ if 
\begin{eqnarray}
\Theta_{12} + \Theta_{3s} < \frac{1}{\sqrt{3}}.
\end{eqnarray}
In particular, it is possible to have
a scalar mass which is arbitrarily large compared to the gaugino mass such that
$m_0 >> m_{1/2}$.  Generically, this may occur if either $\Theta_{12}$ or $\Theta_{3s}$
is negative.

An important question is whether or not this form for the soft terms leads to  
phenomenologically viable superpartner spectra.  It should be noted that these soft terms
correspond to one corner of the of the full mSUGRA/CMSSM parameter space.
A scan of the mSUGRA/CMSSM parameter space was made in~\cite{Mayes:2013qmc} with the trilinear
term fixed as $A_0 = -m_{1/2}$,.  
A plot of this parameter space is shown in Fig.~\ref{fig:mSUGRA_CountourPlanetb30}.
As we can see from this plot, the viable parameter space consist of a strip in the 
$m_0$vs.$m_{1/2}$ plane where $m_0$ is several times larger than $m_{1/2}$.  This, of
course, corresponds to a focus point region of the hyperbolic branch of mSUGRA/CMSSM.
The spectra corresponding to these regions of the parameter space feature squarks and
sleptons with masses above $5$~TeV, a gluino mass in the $3-4$~TeV range, as well as
neutralinos and charginos below $1$~TeV.  The LSP for these spectra is of mixed
bino-higgsino composition with masses in the range $300-800$~GeV.  A plot of the 
direct dark matter detection proton-neutralino cross-sections versus neutralino mass
is shown in Fig.~\ref{fig:DirectDetCrossSectionsvsNeutralinoMasstb30A}.  
As can be seen from this plot, the direct detection cross-sections for these spectra are 
just in the range probed by the XENON100 experiment~\cite{Aprile:2011hi,Aprile:2012nq}.  
In addition, the upcoming XENON1T
experiment~\cite{Aprile:2012zx} will thoroughly cover this parameter space and either will make a discovering
or rule out this parameter space, assuming R-parity conservation, leading to a stable dark matter candidate. 
It should be pointed out that a variation of this model exist where baryon and lepton number
may be gauged, so the imposition of R-parity may not be necessary to solve
the problem of rapid proton decay~\cite{Maxin:2011ne}.

\section{Conclusion}

It has been demonstrated that universal supersymmetry breaking soft terms
may arise in a realistic MSSM constructed in Type II string theory with
intersecting/magnetized D-branes.  In particular, it has been found that these 
soft terms are characterized by a universal scalar mass which is always equal 
to one-half of the gravitino mass, and a universal trilinear term wich is always equal
to the negative of the universal gaugino mass.  For the simplest case where
the goldstino angles for the three complex structure moduli and the dilaton are
all equal, the soft terms are that of the well-known special dilaton.  However,
it was found that more general sets of universal soft terms with different values
for the universal gaugino also exist.  In particular, it was found that it is possible
for the universal scalar mass to be arbitrarily large in comparison to the universal 
gaugino mass.  Thus, for the model which has been under study, it may be natural 
to have scalar masses which are much larger than the gaugino mass.

While the observed mass of the Higgs is below the expected MSSM upper bound, to
obtain a $125$~GeV Higgs mass requires large radiative corrections from the top/stop
sector, implying heavy squarks with multi-TeV masses.  Superpartner spectra with such
large scalar masses may solve the hierarchy problem with low fine-tuning of the 
electroweak scale.  The parameter space corresponding to the particular form of the soft 
terms $m_0 >> m_{1/2}$ and $A_0 = -m_{1/2}$ has been previously studied and results
of this study were reviewed.  Viable spectra from this region of the parameter space
feature squarks and sleptons with masses above $5$~TeV, a $3-4$~TeV gluino mass, as well 
as light neutralinos and charginos at the TeV-scale or below.  In addition, the LSP for
these spectra is of mixed bino-higgsino composition with masses in the range $300-800$~GeV 
and a higgsino fraction of roughly $\approx 70\%$.  
Moreover, the spin-indepenent dark matter direct-detection proton-neutralino cross-sections
are currently being probed by the XENON100 experiment and will be completely tested by the
upcoming XENON1T experiment.  It was shown that the soft terms corresponding to to this 
parameter space naturally and easily obtained from the model.  

\section{Acknowledgments}
The author would like to thank James Maxin and Dimitri Nanopoulos
for helpful discussions while this manuscript was being prepared.

\end{document}